\documentclass[a4paper]{jpconf}
\usepackage[pdftex]{graphicx}
\usepackage[square,sort&compress]{natbib}
\usepackage{url}

\newcommand{\ket}[1]{\left| #1\right\rangle}

\begin{document}
\title{On the spectroscopy of quantum dots in microcavities}

\author{F.P. Laussy and E. del Valle}

 \address{University of Southampton, Highfield, Southampton, SO171BJ, United Kingdom}


\begin{abstract}
  At the occasion of the OECS conference in Madrid, we give a succinct
  account of some recent predictions in the spectroscopy of a quantum
  dot in a microcavity that remain to be observed experimentally,
  sometimes within the reach of the current state of the art.
\end{abstract}

\begin{figure}[hbtp]
  \centering
  \includegraphics[width=\linewidth]{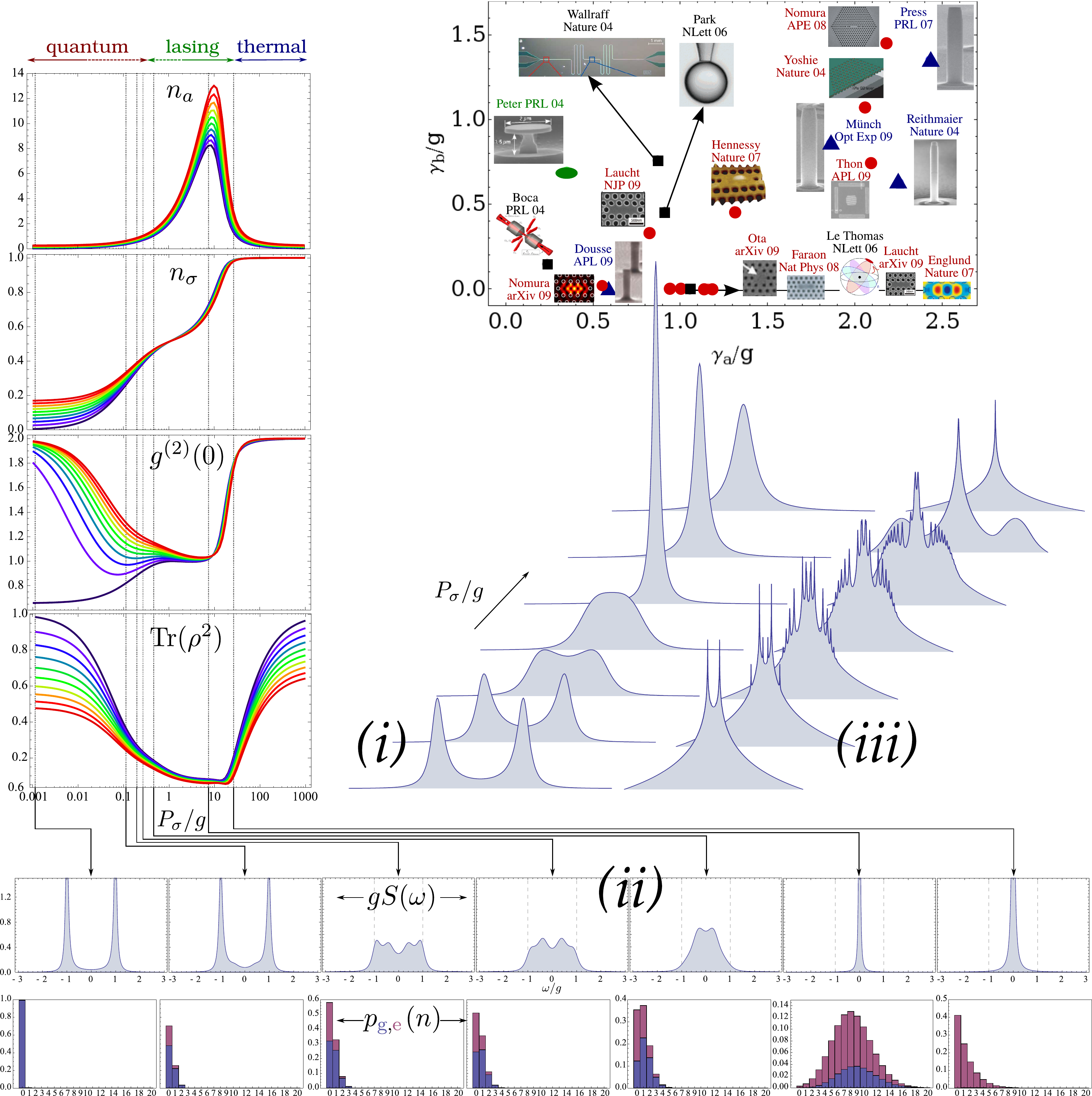}
  \caption{\small\emph{An overall picture of the spectroscopy of
      microcavity quantum electrodynamics}. In inset, positions of
    various systems. Photoluminescence spectra for increasing pumping
    from (for all purposes) vanishing to infinite are shown for, (i),
    a representative of the best currently available systems (as
    reported by Nomura \emph{et al.}~\protect\cite{arXiv_nomura09a}),
    (ii), a realistic near-term device exhibiting explicitly nonlinear
    quantum features, and (iii), an unrealistically good system,
    probing directly the hamiltonian spectrum. Case (i) is in
    agreement with the experimental results. Case
    (iii)---\emph{displayed in log-scale}---exhibits neatly
    transitions from vacuum (Rabi doublet) to quantum (Jaynes-Cummings
    peaks) and lasing (Mollow triplet) before being quenched by
    saturation of the dot. Case (ii) shows the counterpart of such a
    transition in a realistic system, as of today. Quadruplets are
    observed and a transition to lasing follows the emergence of two
    inner peaks, melting into a single line. Below each spectrum, the
    probability $p_{\mathrm{g(e)}}(n)$ of the
    state~$\ket{\mathrm{g(e)},n}$ [for ground(excited) state dressed
    by~$n$ photons] is shown (see also the animation
    \texttt{excitation-statistics.avi}), showing the transition from a
    Fock state (with strong antibuching) to a coherent state and
    ultimately a thermal state for the cavity while the dot is
    maintained in its excited state. Cavity population, $n_a$, quantum
    dot population, $n_\sigma$, $g^{(2)}(0)$ and the trace of $\rho^2$
    are also shown for cavity pumping, from $P_a/g=0$ (blue)
    to 0.05 by steps of 1/180. All spectra are without cavity
    pumping.}
  \label{fig:SatAug29181703BST2009}
\end{figure}

Light-matter coupling of a single quantum dot in a microcavity has
recently enjoyed considerable activity. Led by sustained technological
progress, various groups now control strong coupling~(see
Refs.~\cite{nomura08a,kistner08a,dousse09a,laucht09b} for some recent
reports, and references therein). From a ``naive'' reading of the
figures in the literature, one can establish a rough picture of the
state of the art, positioning various systems (and various groups) in
terms of the cavity, $\gamma_a$, and exciton, $\gamma_b$, decay rates,
in units of the coupling strenght~$g$. This is shown in the inset of
Fig.~\ref{fig:SatAug29181703BST2009}. The closer the system to the
origin, the better it is to exhibit quantum phenomena. We shall
consider in this text (and its supporting media animations) three sets
of parameters:
%
\begin{center}
  (i) $\gamma_a/g=0.55$, $\gamma_\sigma/g=0.014$,\qquad
  (ii) $\gamma_a/g=0.25$, $\gamma_\sigma/g=0$,\qquad 
  (iii) $\gamma_a/g=0.01$, $\gamma_\sigma/g=0$.
\end{center}
The first one is the best system claimed in the literature, by Nomura
\emph{et al.}~\cite{arXiv_nomura09a}, from Arakawa's group. As we
shall shortly discuss, there should be no great confidence entrusted
in the accuracy of our inset, established on the basis of the
estimations of various authors, whose estimations are not always
consistent for a direct comparison nor, in many cases, strictly
consistent in the absolute, for lacking a quantitative analysis.
Point~(ii) is a slightly better system, within the reach of
experiments in the immediate future. Point~(iii) is an unrealistically
good system, considered for its illustration of very marked quantum
features.

We recently studied the regime of strong light-matter coupling between
a two-level system (the quantum dot) and a single cavity mode, under
an incoherent pumping, both for the quantum dot itself, $P_\sigma$,
and for the cavity, $P_a$,~\cite{delvalle09a} extending our treatment
of the linear (and bosonic) regime~\cite{laussy08a}. We summarize and
extend our main findings, and formulate desirable goals (from the
point of view of the authors), both for experimentalists and
for theoreticians.

The investigation itself is of a very fundamental character, since it
assumes a Jaynes-Cummings hamiltonian. In the limit of vanishing
excitation, the case of spontaneous emission is recovered. This case
also matches that of the linear regime, which is also that of coupling
of two boson modes~\cite{laussy09a}, and is therefore solvable
exactly. In this case, peculiarities that can be attributed to the
semiconductor case are the effective quantum state realized in the
system, that can be, beyond the mere excited state of the quantum dot,
an initial state prepared as a photon or, even more likely, an
arbitrary mixed state (a density matrix) of light and matter. This
initial quantum state affects drastically the spectral
shape~\cite{laussy08a,laussy09a}.

When pumping is increased, if the system is indeed described by the
Jaynes-Cummings model, dressed states of the light-matter system are
excited, with potential applications for quantum devices, since they
are ruled by single-photon nonlinearities. As pumping is further
increased, a transition into lasing is observed, that still maintains
strong-coupling, and is therefore closely related to the one-atom
laser. This transition is shown in
Fig.~\ref{fig:SatAug29181703BST2009} for the parameters of Nomura
\emph{et al.} (see also the animation
\texttt{lasing-exciton-pumping.avi}). In qualitative agreement with
the actual experiment, the Rabi doublet collapses, is followed by a
narrow (lasing) line, that is ultimately broadened due to quenching by
the incoherent pumping~\cite{benson99a}. Direct and explicit
manifestation of a few quanta is not directly apparent in case~(i)
with exciton pumping only. Carrying out a similar experiment but
increasing cavity pumping instead, is more prone to give away the
nonlinear features (see the animation
\texttt{lasing-cavity-pumping.avi}). Such nonlinear features, if they
are of a quantum character, should exhibit the characteristic
signature of a dressed states square-root splitting, resulting in
resonances at~$\pm(\sqrt{n}\pm\sqrt{n-1})$ (with~$n$ the number of
excitations in the system). This is strikingly manifest in
system~(iii) (Fig.~\ref{fig:SatAug29181703BST2009}, see also the
animation \texttt{jc-mollow-transitions.avi}). In such a case,
one must observe the system in a suitable rescaled framework (e.g.,
linewidth are vanishing). The results are plotted in log-scale on the
figure (the Rabi doublet sets the scale). A neat transition from the
vacuum Rabi doublet (two peaks) into quenching (Lorentzian line) is
observed. It goes through the quantum regime with multiplets at
anharmonic frequencies, and the classical regime with a Mollow triplet
that results from melting together the distinguishable transitions
between the quantum dressed states. For the case of point~(ii), we
provide a more detailed picture by supplementing the spectral shape
evolution, with the main observables of interest (see also the
animation \texttt{realistic-jc-structure.avi}). These are the cavity
population~$n_a$, the probability of the quantum dot to be in its
excited state~$n_\sigma$ and the two-photon counting probability at
zero delay~$g^{(2)}(0)$. We also plot the trace of $\rho^2$ ($\rho$
being the total density matrix), that gives the order of coherence (or
purity) of the system, and the probabilities $p_{g(e)}(n)$ of the
system to be found with $n$ photons dressing the ground (excited)
state (stacked on top of each other in the figure, so that the total
height is their sum, that represents the cavity photon statistics).

\section*{Challenges for experimentalists and theoreticians}

We list a short series of goals that we regard as open, realistic
given currently available systems and important results, whether
obtained or infirmed. The first three ``challenges'' are experimental
and the last two theoretical:

\begin{enumerate}
\item Alter the effective quantum state realized in the system.\label{c1}
\item Compare cavity and direct exciton emission.\label{c2}
\item Evidence Jaynes-Cummings features in the spectral shape.\label{c3}
\item Solve exactly the Jaynes-Cummings model with incoherent pumping.\label{c4}
\item Provide a statistical description of data.\label{c5}
\end{enumerate}

(i) We already commented on the sensibility of the spectral shape on
the effective quantum state of the system. Beside, the lasing of the
strongly-coupled light-matter system bears some differences whether
driven by the exciton pumping or by the cavity pumping (that we shall
discuss somewhere else). For these two reasons, it is a desirable goal
to be able to control the photon-like or exciton-like character of the
state, ideally up to an arbitrary ratio of the photon versus the
exciton component (both in the spontaneous emission and nonlinear
regimes). A striking manifestation of this effect would be to unravel
the inner Jaynes-Cummings peaks by bringing the system from exciton-like
to photon-like.

(ii) In the linear regime, there is a symmetry between the effective
quantum state and the channel of emission (through the cavity mode or
the direct exciton emission). In the nonlinear regime, such an
equivalence breaks down, and is replaced by richer spectroscopic
structures. In particular, the direct exciton emission is usually more
prone to exhibit quantum phenomena, in particular the Jaynes-Cummings
ladder or the Mollow triplet (that is seen on
Fig.~\ref{fig:SatAug29181703BST2009} mainly because for such
exceedingly good systems, one probes directly the hamiltonian spectrum
rather than the cavity or the exciton optical spectra). For this
reason, it is desirable to be able to compare spectral shapes of both
the cavity and direct exciton emission.

(iii) Obviously, a direct observation of the nonlinear Jaynes-Cummings
structure would be an important achievement. It can be helped relying
on the two previous points, or it could be probed by resonant
excitation. If the coupling is dephased at a similar rate as the
coherent light-matter coupling, the characteristic Jaynes-Cummings
multiplet gives rise to a
triplet~\cite{arXiv_gonzaleztudela09a}. Maybe Jaynes-Cummings
nonlinearities have been already observed, in this
disguise~\cite{hennessy07a,arXiv_ota09a}.

(iv) One of the interesting feature of the Jaynes-Cummings model is
that it admits analytical solutions. To the best of our knowledge,
none is known yet in the presence of an incoherent pumping (results in
Fig.~\ref{fig:SatAug29181703BST2009} are obtained numerically).
Beside the interest for its own sake, the exact solution would allow
convenient optimization, e.g., to obtain the best antibunching to
signal ratio.  It would also explain features such as the existence of
a fixed point for~$n_\sigma(P_a,P_\sigma)$, a possible criterion to
define lasing in such systems. Last but not least, if solutions can be
found for two-time correlators, it would greatly help in fitting
experimental data.

(v) So far, most reports have been based on qualitative features, such
as observation of an anticrossing (but see Refs.~\cite{laucht09b}
and~\cite{munch09a} for some notable, and laudable,
exceptions). Beside the possibility to be even qualitatively
incorrect, this approach has the serious shortcoming of great
inaccuracy, that impedes understanding and biases estimates of the
systems and how far they are of reaching predicted features. We expect
that the inset of Fig.~\ref{fig:SatAug29181703BST2009} would change
significantly if positioning was made after statistical inference
rather than reading of linewidths in a Lorentzian fitting, or fitting
of maxima (which is impossible~\cite{arXiv_gonzalez-tudela09b}). A
statistical analysis of the data, providing intervals of confidences
for the fitting parameters, is not trivial in the nonlinear case,
where conventional nonlinear regression methods do not apply, for
lacking a closed-solution of the model. This would however demonstrate
the possibility of such systems to be accounted for exactly (within
standard deviates), and entitle the community to speak of
\emph{microcavity quantum electrodynamics}.

\small
\bibliographystyle{iopart-num}
\bibliography{Sci,arXiv}

\vfill\eject

\ttfamily
The animations mentioned in this text are available at the following adresses:
\begin{enumerate}
  \setlength{\itemsep}{0pt}
\item http://laussy.org/wp-content/uploads/2009/09/lasing-exciton-pumping.avi
\item http://laussy.org/wp-content/uploads/2009/09/lasing-cavity-pumping.avi
\item http://laussy.org/wp-content/uploads/2009/09/jc-mollow-transition.avi
\item http://laussy.org/wp-content/uploads/2009/09/realistic-jc-structure.avi
\item http://laussy.org/wp-content/uploads/2009/09/excitation-statistics.avi
\end{enumerate}

\end{document}